\documentclass[12pt]{article}
\usepackage{epsf}
\usepackage[dvips]{color}
\usepackage{exscale}
\usepackage[dvips]{graphicx}
\usepackage{amsmath}
\usepackage{amssymb}
\usepackage{latexsym}

\begin{document}
\setlength{\baselineskip}{18pt}
\begin{titlepage}
\begin{flushright}
KUNS-2274\\
YITP-10-47
\end{flushright}

\vspace*{1.2cm}
\begin{center}
{\Large\bf Revisiting superparticle spectra \\ in superconformal 
flavor models}
\end{center}
\lineskip .75em
\vskip 1.5cm

\begin{center}
{\large 
Tatsuo Kobayashi$^{a,}$\footnote[1]{E-mail:
\tt kobayash@gauge.scphys.kyoto-u.ac.jp}, 
Yuichiro Nakai$^{b,}$\footnote[2]{E-mail:
\tt ynakai@yukawa.kyoto-u.ac.jp}, 
and 
Ryo Takahashi$^{c,}$\footnote[3]{E-mail: 
\tt ryo.takahashi@mpi-hd.mpg.de}
}\\

\vspace{1cm}

$^a${\it Department of Physics, Kyoto University, Kyoto 606-8502, Japan}\\
$^b${\it Yukawa Institute for Theoretical Physics, Kyoto University, Kyoto 
606-8502, Japan}
$^c${\it Max-Planck-Institut f$\ddot{u}$r Kernphysik, Postfach 10 39 80, 69029
Heidelberg, Germany}\\

\vspace*{10mm}
{\bf Abstract}\\[5mm]
{\parbox{13cm}{\hspace{5mm}
We study superparticle spectra in the superconformal flavor scenario
with non-universal gaugino masses. The non-universality of gaugino
masses can lead to the wino-like or higgsino-like neutralino
LSP. Furthermore, it is shown that the parameter space for the
higgsino-like LSP includes the region where the fine-tuning problem
can be improved. 
The degeneracy of soft scalar masses does not
drastically change when ratios of gaugino masses are of ${\cal O}(1)$.
The degeneracy of scalar masses for squarks and left-handed
sleptons would be good enough to avoid the FCNC problem but that of
right-handed slepton masses is weak. 
However, the overall size of right-handed slepton masses becomes
larger when the bino becomes heavier. 
Thus, that would be favorable to avoid the FCNC problem 
due to soft scalar masses and A-terms.
}}
\end{center}
\end{titlepage}

\section{Introduction}

Most of free parameters in the standard model appear in 
the Yukawa couplings leading to quark/lepton masses 
and mixing angle.
It is important to understand 
what is the origin of hierarchical Yukawa couplings.
Indeed, various flavor mechanisms have been proposed 
to derive the hierarchy of Yukawa couplings.

Supersymmetric extension of the standard model 
is one of interesting candidates as physics beyond
the standard model.
Flavor mechanisms leading to hierarchical Yukawa couplings
would also affect superpartners of quarks and leptons.
Then, we would have a prediction on squark and slepton masses 
in each flavor mechanism.
Such a prediction could be tested when squarks and sleptons 
are discovered.
Furthermore, we have already experimental constraints on 
squark and slepton masses by flavor changing neutral 
current (FCNC) processes, even though squarks and sleptons have 
not been discovered yet~\cite{FCNCbound}.\footnote{See also 
e.g  Ref.~\cite{Chankowski:2005jh,Altmannshofer:2009ne} 
and references therein.} 
Such constraints require a certain degree of degeneracy 
among squark and slepton masses when those masses are 
of ${\cal O}(100)$ GeV.
In addition, A-terms can be another source for the FCNC problem.
Then, alignment between Yukawa matrices and A-term matrices 
is required when squark and slepton masses are of ${\cal O}(100)$
GeV.

The flavor mechanism proposed by Nelson and Strassler
\cite{Nelson:2000sn} is one of interesting mechanisms 
to lead to hierarchical Yukawa couplings.
In this flavor scenario,
the supersymmetric standard model (SSM) couples with a 
superconformal (SC) sector between certain energy scales, 
and it decouples below the energy scale $M_c$.
Then, superconformal dynamics leads to suppressed Yukawa couplings 
of quarks and leptons
at low energy even if those values are of ${\cal O}(1)$ 
at high energy.
(See for early works on this scenario 
Ref.~\cite{Kobayashi:2001kz,Nelson:2001mq,Kobayashi:2001is}.
Also, new models were proposed recently 
\cite{Poland:2009yb,Aharony:2010ch,Craig:2010ip,Abel:2010kw}.)
Furthermore, superconformal dynamics has another interesting 
aspect.
Suppose that supersymmetry (SUSY) is broken softly 
at a high energy scale, which is higher than $M_c$,
and generic values of soft SUSY breaking terms appear in 
the SSM sector as well as the SC sector.
However, superconformal dynamics can suppress 
squark and slepton masses for any initial values of 
sfermion masses \cite{Karch:1998qa,Kobayashi:2001kz,Nelson:2001mq}.
Then, small values of squark and slepton masses 
could be realized at the decoupling energy scale $M_c$.
Below the decoupling scale $M_c$, squark and slepton masses 
receive usual radiative corrections due to the SSM gaugino masses.
Since such radiative corrections are flavor-blind and sizable, 
we could realize almost degenerate squark and slepton masses 
at the weak scale.
That is, the superconformal dynamics could wash out 
initial conditions of squark and slepton masses 
at least for the first and second generations and 
their degeneracy could be realized at the weak scale.
This soft mass spectrum is similar to that in the gaugino 
mediation~\cite{Kaplan:1999ac}.
This superconformal flavor mechanism is quite interesting from the viewpoints 
of both generation of hierarchical Yukawa couplings 
and dynamical realization of degenerate sfermion masses.
We do not need to require a specific SUSY breaking mediation 
mechanism to avoid the SUSY FCNC problem.

Within this scenario, detailed studies on the low energy spectrum 
have been done in Ref.~\cite{Kobayashi:2001kz,Nelson:2001mq} 
by assuming the universal gaugino mass at the energy scale of 
grand unified theory (GUT), i.e. the GUT relation.
Then, it is found that the right-handed sleptons tend to become 
the lightest superparticle (LSP), unless the gravitino is 
lighter than them.
The slepton LSP is unfavorable from the cosmological viewpoint.
In addition, it is found that sfermion masses are 
expected to be degenerate.
However, 
right-handed slepton masses differ from each other by ${\cal O}(10)
\%$, that is, their degeneracy is rather weak.
For the other sfermion masses, 
 their degeneracy is 
much better like ${\cal O}(1)\%$ -- ${\cal O}(0.1)\%$.
The $U(1)_Y$ D-term, $S={\rm Tr}Ym^2_i$, is helpful both to 
make the right-handed slepton heavier than the bino and to 
improve the degeneracy between right-handed slepton masses.
However, we may need a sizable value of the $U(1)_Y$ D-term.
In general, the ratios of A-term matrices to Yukawa matrices 
can not be controlled by the superconformal dynamics.
That may lead to the FCNC problem when 
squark and slepton masses are of ${\cal O}(100)$ GeV.

In this article, we study the superparticle spectra by 
relaxing the GUT relation among the SSM gaugino masses.
Indeed, several mechanisms lead to non-universal gaugino 
masses \cite{Choi:2007ka}, 
e.g. moduli mediation \cite{Brignole:1993dj}, 
gravity mediation by F-terms of gauge non-singlets \cite{Ellis:1984bm},
anomaly mediation \cite{Randall:1998uk}, 
gauge messenger scenario \cite{Dermisek:2006qj},
general gauge mediation \cite{Meade:2008wd}, 
mirage mediation \cite{Choi:2004sx,Choi:2005uz}, etc.
The reason of the slepton LSP in the above analysis is that 
radiative corrections due to the bino mass are not large.\footnote{
Also, see e.g. \cite{Komine:2000tj,Baer:2002by}.}
Thus, relaxing the GUT relation, in particular a large 
bino mass, would be helpful to avoid the slepton LSP.
Furthermore, a certain ratio of non-universal gaugino masses 
is useful to improve/ameliorate the fine-tuning problem 
\cite{Choi:2005hd,Abe:2007kf,Gogoladze:2009bd,Horton:2009ed,Kobayashi:2009rn}.
It is also important to study the degeneracy among sfermion masses 
under the assumption of non-universal gaugino masses.
These are our purposes in this paper.

This paper is organized as follows.
In section 2, we briefly review the superconformal 
flavor scenario and give values of sfermion masses 
at $M_c$.
In section 3, by assuming the non-universal gaugino masses, 
we study the superparticle spectra, 
in particular, the LSP, and also investigate 
the fine-tuning of the Higgs sector.
In addition, we study the degeneracy among 
squark and slepton masses.
We also discuss about A-terms.
Section 4 is devoted to conclusion and discussion.

\section{Superconformal flavor scenario}

In this section, we briefly review the superconformal 
flavor scenario \cite{Nelson:2000sn}, 
in particular values of squark and slepton masses 
at the decoupling scale $M_c$ \cite{Kobayashi:2001kz,Nelson:2001mq}.
We assume two sectors.
One is the SSM sector, which has 
the gauge group $SU(3) \times SU(2) \times U(1)_Y$, three 
families of matter fields and a pair of Higgs fields, $H_u$ and $H_d$.
It is straightforward to extend it such as GUT groups and 
more matter/Higgs fields.
The other sector is the SC sector including 
the gauge group $G_{SC}$ and matter fields $\Phi_r$.
We consider the following superpotential,
\begin{eqnarray}\label{eq:super-p}
W=y^{ij}q_iq_jH + \lambda^{rsi}\Phi_r\Phi_s
q_i + \hat W(\Phi),
\end{eqnarray} 
where $q_{i,j}$ denote quarks and leptons in the SSM 
and $H$ corresponds to $H_u$ and $H_d$.
Here, the first term denotes the ordinary Yukawa couplings 
among quarks/leptons and Higgs fields.
On the other hand, the second term corresponds to 
couplings between the SSM matter fields $q_i$ 
and the SC matter fields $\Phi_{r,s}$.
Some of SC matter fields must have non-trivial representations 
under the standard-model gauge group in order to allow these 
couplings.
The third term denotes couplings among only the SC matter fields.

We assume that the superconformal dynamics generates 
sizable and negative anomalous dimensions of 
the SC matter fields, $\Phi_r$, in the conformal regime 
between $\Lambda$ and $M_c$, where $\Lambda > M_c$.
Also we assume that the coupling $\lambda^{rsi}$ approaches 
toward a fixed point.
Then, sizable and positive anomalous dimensions $\gamma_i$ of 
the SSM matter fields $q_i$ are generated.
As a result, the Yukawa couplings $y^{ij}$ at $M_c$ are 
obtained as 
\begin{eqnarray}\label{eq:yukawa}
y^{ij}(M_c) = y^{ij}(\Lambda) \left( \frac{M_c}{\Lambda}
\right)^{(\gamma_i +\gamma_j)/2}.
\end{eqnarray}
Even if $y^{ij}(\Lambda) ={\cal O}(1)$, 
we would realize suppressed Yukawa couplings $y^{ij}(M_c)$ at 
$M_c$, depending on values of anomalous dimensions and 
the length of conformal regime, i.e. $M_c/\Lambda$.
Thus, this scenario is interesting to generate 
hierarchical Yukawa couplings among three families.
In order to derive the realistic fermion mass hierarchy, 
one needs different anomalous dimensions for three 
families, when the length of conformal regime is the same 
between different families.
The third family may not couple to the SC sector in order not to 
suppress the Yukawa coupling at least for the top quark.

In addition to the above generation of Yukawa hierarchy, 
the superconformal dynamics has another important aspect.
Suppose that the SUSY is broken softly above $\Lambda$ 
and soft SUSY breaking terms are induced.
The SC sector suppresses the gaugino mass 
of the SC sector, soft scalar masses of $\Phi_r$  
and squark/slepton masses of $q_i$, which couple to 
the SC sector \cite{Karch:1998qa,Kobayashi:2001kz,Nelson:2001mq}.
However, the superconformal symmetry is not exact, but 
it is broken by the gauge couplings $g_a$ and gaugino masses $M_a$ 
($a=1,2,3$) in the SSM sector.
Then, squark/slepton masses $m_{\tilde q_i}$ of $q_i$ do not continue to 
suppress, but converge to definite values at $M_c$
\cite{Kobayashi:2001kz,Nelson:2001mq},
\begin{eqnarray}\label{eq:m-converge}
m_{\tilde q_i}^2 \rightarrow 
\sum_a\frac{C_{ia}}{\Gamma_i}\alpha_a(M_c)M_a^2(M_c), 
\end{eqnarray}
where $\alpha_a=g_a^2/(8 \pi^2)$ and $C_{ia}$ denotes the quadratic 
Casimir.
Here, $\Gamma_i$ is a numerical factor of ${\cal O}(1)$, 
which is obtained from the anomalous dimensions, and therefore 
$\Gamma_i$ is flavor-dependent.
For concreteness, we show the above convergent values 
for all squark/slepton masses,
\begin{eqnarray}\label{eq:m-converge-2}
m_{\tilde Qi}^2(M_c) &=& {1 \over \Gamma_{Qi}}
[{16 \over 3}\alpha_3M_3^2 + 3\alpha_2M_2^2 
+{1 \over 15}\alpha_1M_1^2 ](M_c), \nonumber \\
m_{\tilde ui}^2(M_c) &=& {1 \over \Gamma_{ui}}
[{16 \over 3}\alpha_3M_3^2 + {16 \over 15}\alpha_1M_1^2 ](M_c), 
\nonumber \\
m_{\tilde di}^2(M_c) &=& {1 \over \Gamma_{di}}
[{16 \over 3}\alpha_3M_3^2 + {4 \over 15}\alpha_1M_1^2 ](M_c), \\
m_{\tilde Li}^2(M_c) &=& {1 \over \Gamma_{Li}}
[3\alpha_2M_2^2 +{3 \over 5}\alpha_1M_1^2 ](M_c), \nonumber \\
m_{\tilde ei}^2(M_c) &=& {1 \over \Gamma_{ei}}
[{12 \over 5}\alpha_1M_1^2 ](M_c),   \nonumber
\end{eqnarray}
where $Q_i, u_i, d_i, L_i$ and $e_i$ denote 
left-handed quarks, right-handed up-sector quarks, 
right-handed down-sector quarks, 
left-handed leptons and right-handed leptons, 
respectively.

The above values of squark/slepton masses squared at $M_c$ are 
smaller by a loop factor $\alpha_a$ than $M_a^2$, although 
those squarks/slepton masses squared are flavor-dependent.
After the decoupling of the SC sector at $M_c$, 
squark/slepton masses squared $m_{q_i}^2$ receive flavor-independent 
radiative corrections 
due to the SSM gaugino masses $M_a$ between $M_c$ and the weak scale.
Such radiative corrections are sizable and of ${\cal O}(M_a^2)$ 
when $M_c$ is sufficiently higher than the weak scale $M_Z$.
In this case, the overall size of squark/slepton masses at $M_Z$ 
is determined by radiative corrections between $M_c$ and $M_Z$.

It is found that the right-handed slepton mass $m_{\tilde e}$ is smaller than 
the bino mass $M_1$, i.e.
\begin{eqnarray} \label{eq:slepton-LSP}
m_{\tilde e}(M_Z) < M_1(M_Z),
\end{eqnarray}
 when $M_c \leq 2 \times 10^{16}$ GeV.
Then, the right-handed slepton would be the LSP 
\cite{Kobayashi:2001kz,Nelson:2001mq} if 
we assume the GUT relation of gaugino masses,
\begin{eqnarray} \label{eq:GUT-relation}
M_3(M_c) : M_2(M_c) : M_1(M_c) = \alpha_3 (M_c) :  \alpha_2 (M_c) :  
 \alpha_1 (M_c), 
\end{eqnarray}
 and the gravitino is heavier.
That is unfavorable from the cosmological viewpoint.

{}From the viewpoint of the SUSY flavor problem, 
it is quite important that squark and slepton masses 
at $M_c$ are suppressed by a loop factor compared with 
gaugino masses $M_a$, although those squark and slepton masses 
are flavor-dependent at $M_c$.
As said above, flavor-independent radiative corrections 
between $M_c$ and $M_Z$ are dominant in the overall size 
of squark and slepton masses.
Hence, squark and slepton masses are almost degenerate.
The difference of those masses squared between different families 
is evaluated as 
\begin{eqnarray}
m_{q_i}^2(M_Z)-m_{q_j}^2(M_Z) = 
m_{q_i}^2(M_c)-m_{q_j}^2(M_c) = \left( \frac{1}{\Gamma_i} -
\frac{1}{\Gamma_j} \right)
\sum_a {C_{ia}}\alpha_a(M_c)M_a^2(M_c).
\end{eqnarray}
When the GUT relation (\ref{eq:GUT-relation}) is assumed, 
the degeneracy between soft masses is better in the squark sector 
and there would be no problem in FCNC experiments.
The degeneracy is weak in the slepton sector, 
in particular for the right-handed sleptons.
The difference of right-handed slepton masses between different 
families is of ${\cal O}(10)\%$.
Such a weak degeneracy may be problematic when 
the slepton mass is of ${\cal O}(100)$ GeV.

The above aspects such as the LSP and mass degeneracy 
depend on gaugino masses.
Indeed, the reason of the slepton LSP is the GUT relation 
(\ref{eq:GUT-relation}), unless the $U(1)_Y$ D-term is sizable.
If we assume other values of gaugino mass ratios, 
we would have different aspects.
That is what we will study in the next section.

\section{Non-universal gaugino masses}

Several mechanisms lead to non-universal gaugino masses.
Thus, we relax the relation (\ref{eq:GUT-relation}) 
and here we consider the following ratio of gaugino masses, 
\begin{eqnarray}\label{eq:non-universal}
M_3(M_c) : M_2(M_c) : M_1(M_c) = \alpha_3 (M_c) :  r_2 \alpha_2 (M_c) :  
r_1 \alpha_1 (M_c), 
\end{eqnarray}
where $r_1$ and $r_2$ are real parameters.
Each SUSY breaking mediation mechanism would lead to 
a certain ratio, $(r_1,r_2)$.
With the above relation (\ref{eq:non-universal}), 
we study phenomenological aspects such as the LSP and 
mass degeneracy by varying the parameters $(r_1,r_2)$.
In the following analysis, we assume that the $U(1)_Y$ 
D-term is not sizable and neglect it.

\subsection{Superparticle spectrum}

Here, we study the superparticle spectra, in particular
the LSP.
The right-handed slepton masses receive radiative corrections
only from the bino mass.
Thus, even if we consider the non-universal relation
 (\ref{eq:non-universal}), the right-handed slepton is 
lighter than the bino for $M_c \leq 2 \times 10^{16}$ GeV, 
i.e. (\ref{eq:slepton-LSP}).
This implies that the bino can not become the LSP, 
while other superparticles are candidates for the LSP.
Thus, only the wino-like or higgsino-like neutralino 
can be the LSP among neutralinos.
We use the following relation,
\begin{eqnarray}\label{eq:mu-MZ}
\mu^2(M_Z) = -\frac{M_Z^2}{2}- \frac{m^2_{Hu}(M_Z) \tan^2 \beta - 
m^2_{Hd}(M_Z) }{\tan^2 \beta -1},
\end{eqnarray}
in order to evaluate the higgsino mass $\mu(M_Z)$.
Here, $m_{Hu}(M_Z)$ and $m_{Hd}(M_Z)$ denote soft scalar 
masses of the up and down Higgs fields, $H_u$ and $H_d$, 
at $M_Z$.
To evaluate  $m^2_{Hu}(M_Z)$ and $m^2_{Hd}(M_Z)$, 
one needs their initial values at $M_c$ and 
stop masses, $m_{\tilde Q_3}$ and $m_{\tilde u_3}$ and 
the A-term $A_t$ corresponding to the top Yukawa coupling, 
because these as well as gaugino masses $M_a$ 
contribute to radiative corrections of  $m^2_{Hu}(M_Z)$ and
$m^2_{Hd}(M_Z)$.
The third family and the Higgs fields do not (strongly) couple to 
the SC sector in order to lead to the large top Yukawa coupling.
Thus, there is no prediction from the superconformal dynamics 
like Eqs.~(\ref{eq:m-converge}) and (\ref{eq:m-converge-2}) for 
the first and second generations.
Here, let us consider the following initial values,
\begin{eqnarray}
&{\rm (i)}& ~~~m_{Q3}=m_{u3}=m_{Hu}=m_{Hd}=A_t=0, \\
&{\rm (ii)}& ~~~ m_{Q3}=m_{u3}=m_{Hu}=m_{Hd}=A_t=M_3, \\
&{\rm (iii)}& ~~~ m_{Q3}=m_{u3}=m_{Hu}=m_{Hd}=3M_3, \quad A_t=0, 
\end{eqnarray}
at $M_c$, 
for simplicity.
Then, by varying $(r_1,r_2)$ we can evaluate 
the superparticle spectra for the cases (i), (ii) and (iii).

\begin{figure}
\begin{center}
\begin{tabular}{cc}
(a) $M_c=2\times10^{16}$ GeV & (b) $M_c=10^{14}$ GeV \\
\includegraphics[scale = 0.92]{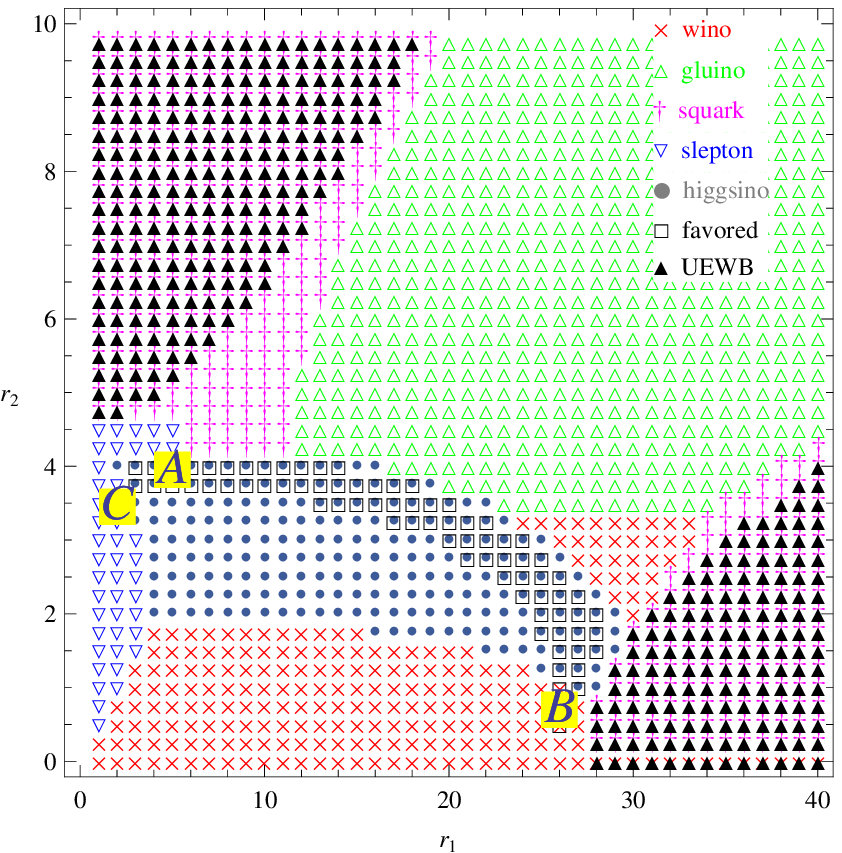} &
\includegraphics[scale = 0.92]{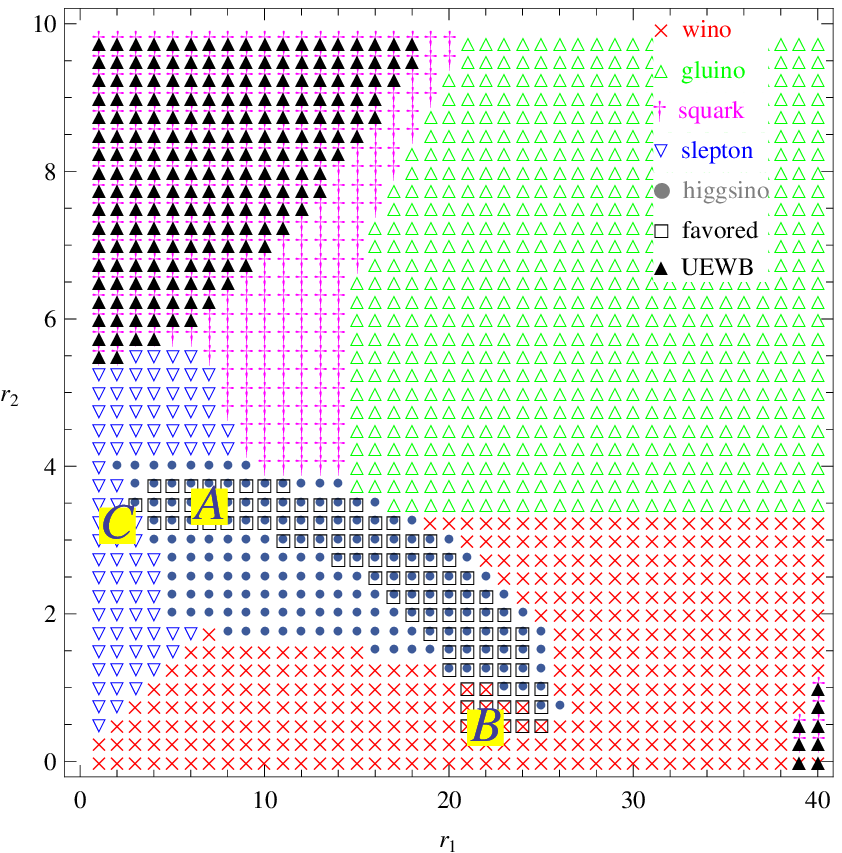} 
\end{tabular}

(c) $M_c=10^{12}$ GeV

\includegraphics[scale = 0.92]{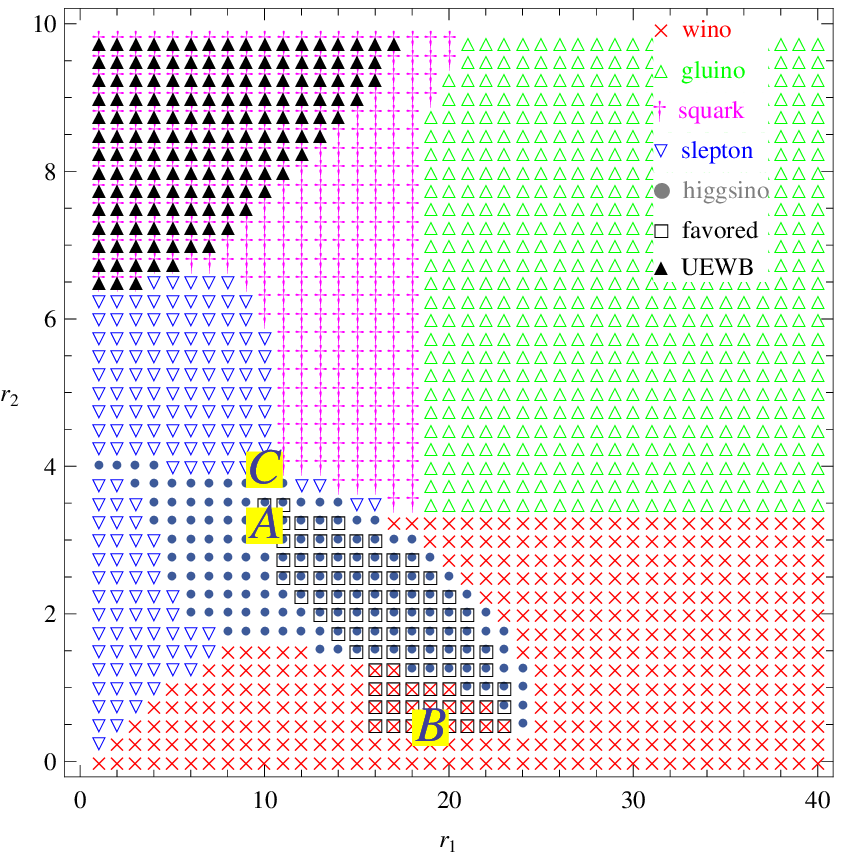}
\end{center}
\caption{LSP in the case (i): The favored region described by the square means that the fine-tuning problem can be relatively relaxed, $\Delta_\mu\lesssim20$.}
\label{fig1}
\end{figure}

\begin{figure}
\begin{center}
\begin{tabular}{cc}
(a) $M_c=2\times10^{16}$ GeV & (b) $M_c=10^{14}$ GeV \\
\includegraphics[scale = 0.92]{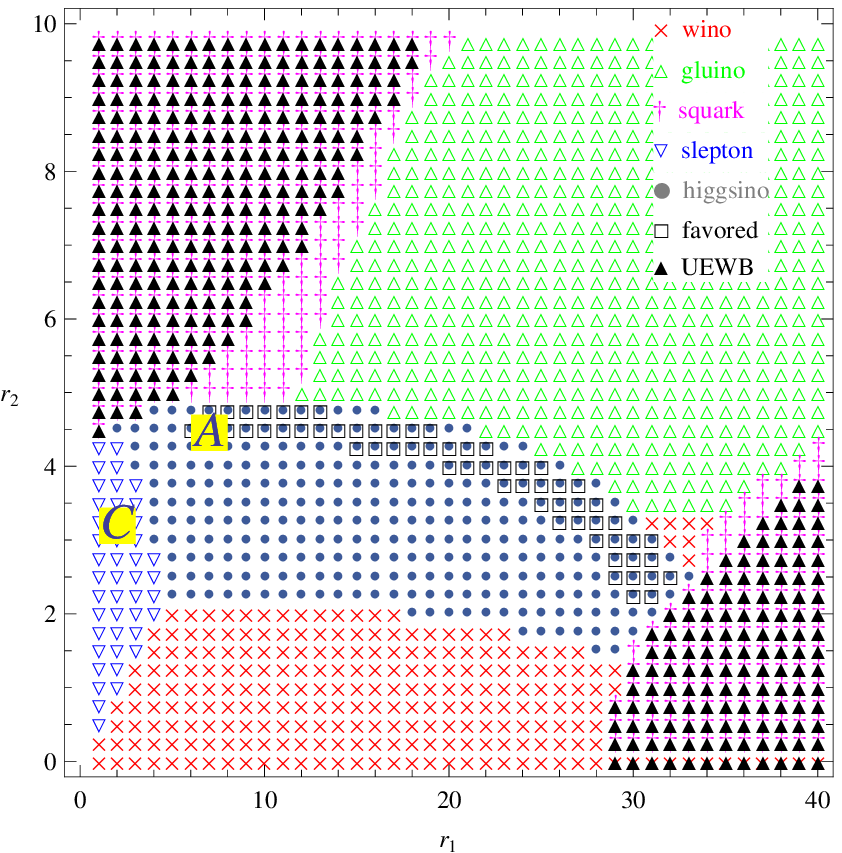} &
\includegraphics[scale = 0.92]{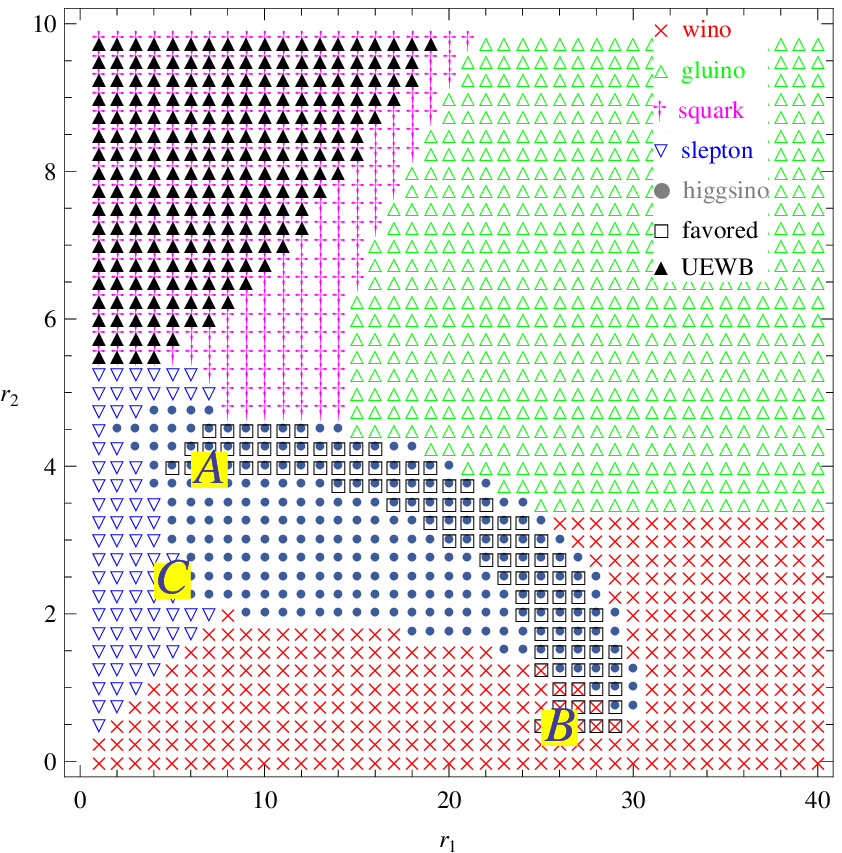} 
\end{tabular}

(c) $M_c=10^{12}$ GeV

\includegraphics[scale = 0.92]{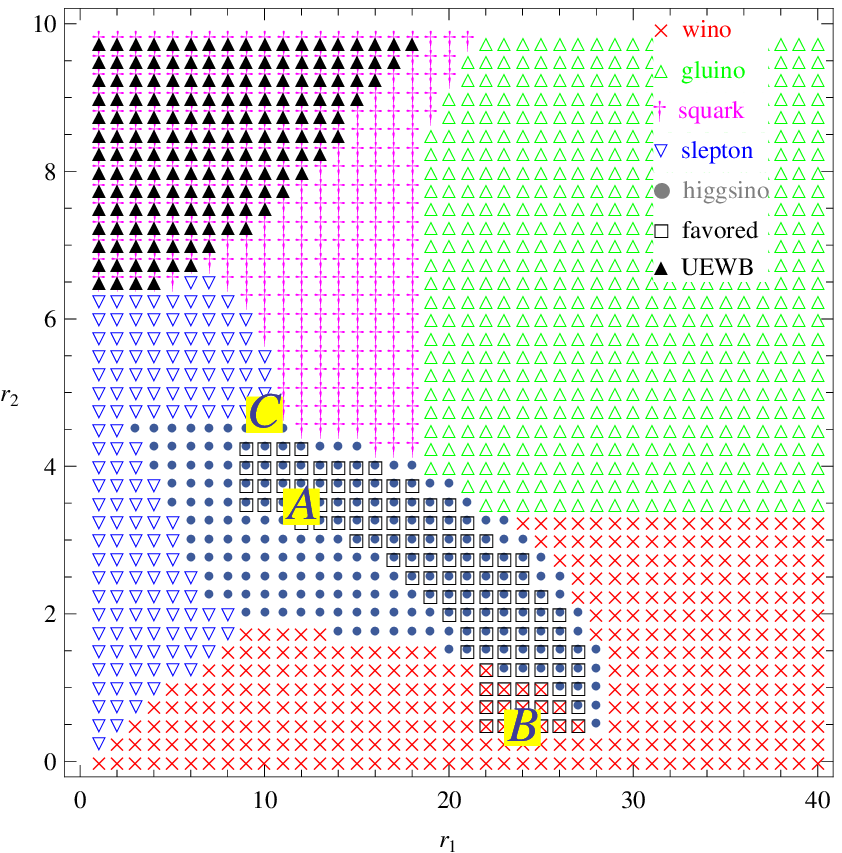}
\end{center}
\caption{LSP in the case (ii): The favored region described by the square means that the fine-tuning problem can be relatively relaxed, $\Delta_\mu\lesssim20$.}
\label{fig2}
\end{figure}

\begin{figure}
\begin{center}
\begin{tabular}{cc}
(a) $M_c=2\times10^{16}$ GeV & (b) $M_c=10^{14}$ GeV \\
\includegraphics[scale = 0.92]{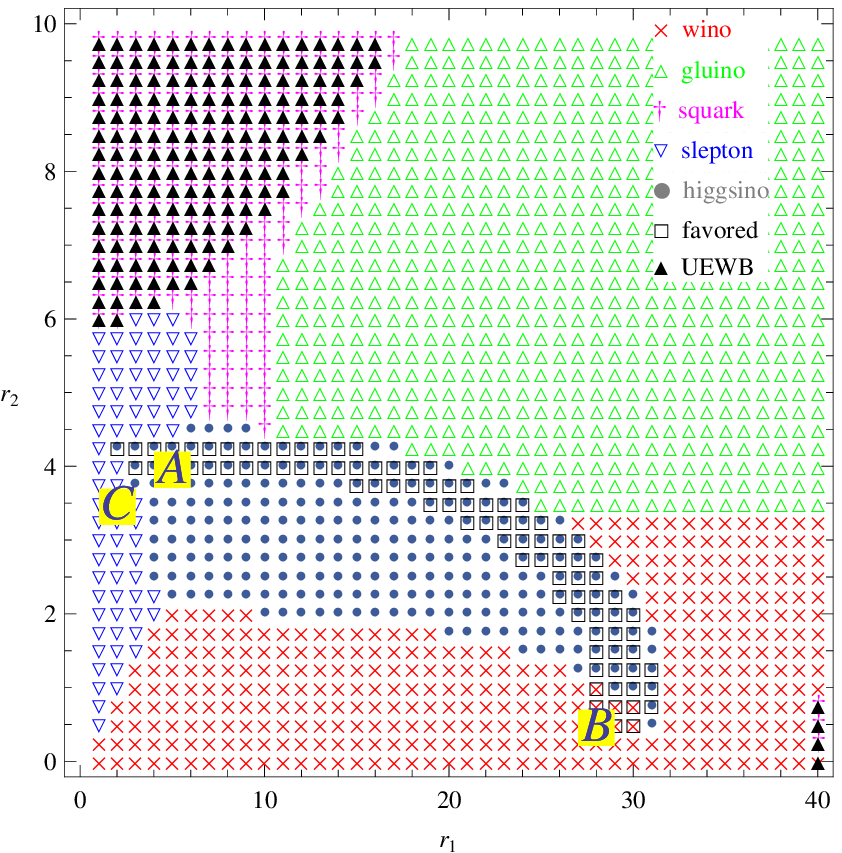} &
\includegraphics[scale = 0.92]{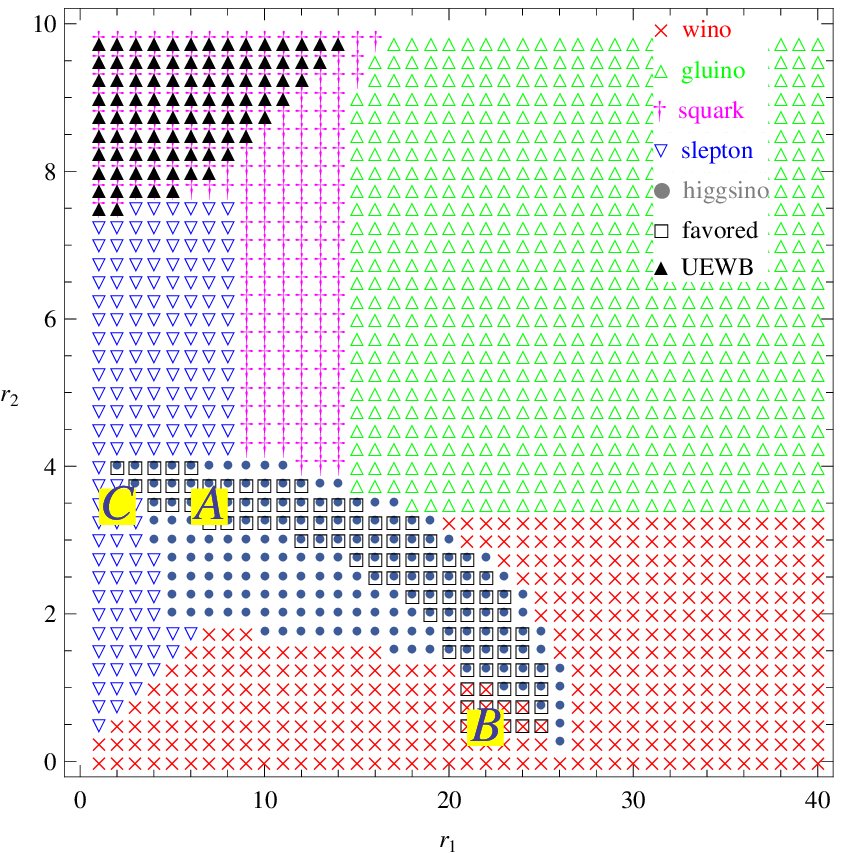} 
\end{tabular}

(c) $M_c=10^{12}$ GeV

\includegraphics[scale = 0.92]{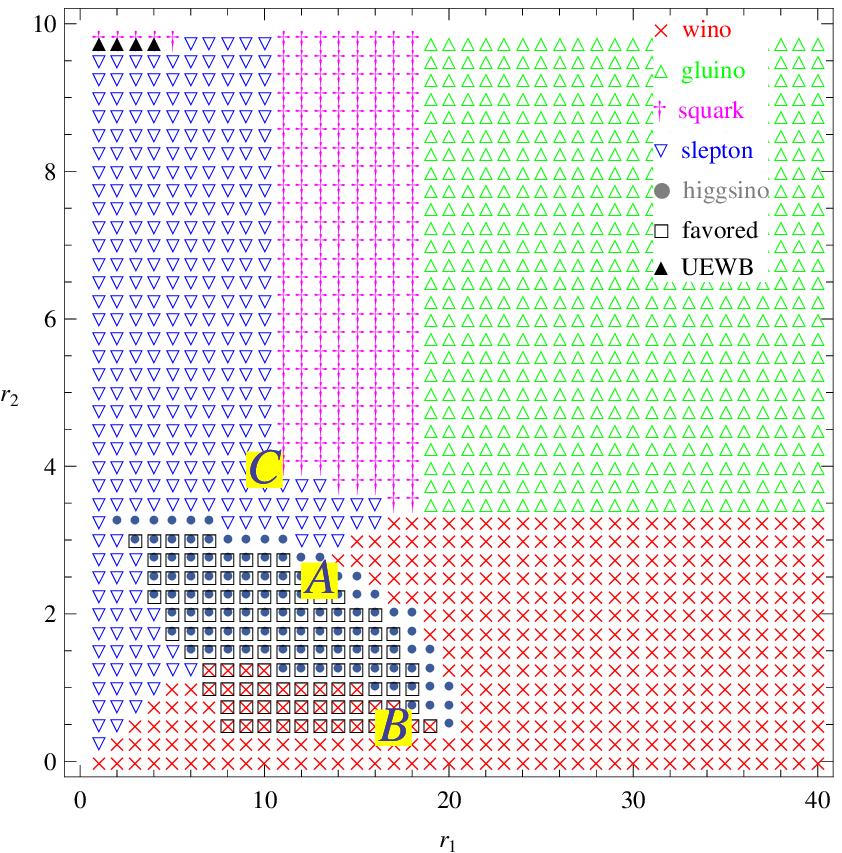}
\end{center}
\caption{LSP in the case (iii): The favored region described by the square means that the fine-tuning problem can be relatively relaxed, $\Delta_\mu\lesssim20$.}
\label{fig3}
\end{figure}

For example, we take $M_3(M_c)=300$ GeV and $\tan \beta =10$.
Figures \ref{fig1} (a)$-$(c)  show the LSP in the case (i) 
for $M_c=2 \times 10^{16}$, $10^{14}$, and $10^{12}$ GeV, respectively.
Figures \ref{fig2} (a)$-$(c) and \ref{fig3}  (a)$-$(c) are the same figures 
for the cases (ii) and (iii), respectively.

When we vary $(r_1,r_2)$, we can not realize successfully radiative 
electroweak symmetry breaking (EWB) in a certain region, where 
the stop becomes tachyonic instead of the Higgs scalar.
Such regions with unsuccessful EWB (UEWB) are shown by black triangles 
in the figures.
Around such regions, there is the parameter space corresponding to 
the squark LSP.
The GUT relation, i.e. $(r_1,r_2)=(1,1)$, leads to 
the slepton LSP and 
there is also the region with small $r_1$, which  
leads to the slepton LSP.
The parameter region with large values of both $r_1$ and $r_2$ 
leads to the gluino LSP. 
The regions with (red) ``$\times$'' including the region with squares, 
  correspond to the parameter space, where 
the LSP is wino-like.
The regions with (gray) dots including the region with squares, 
correspond to the parameter space, where 
the LSP is higgsino-like.  
Later, we will explain about the region with squares.
It is remarkable that there exist broad regions giving the wino- or
higgsino-like LSP as long as the value of $r_2$ is of ${\cal O}(1)$
such as 
$r_2\lesssim4$. The wino-like neutralino tends to be the LSP around
relatively large value of $r_1$ such as $r_1=\mathcal{O}(30 - 40)$ 
with $r_2={\cal O}(1)$.

The light higgsino would be favorable to avoid 
the fine-tuning problem.
That is, from the relation (\ref{eq:mu-MZ}), it 
would be natural that $\mu(M_Z)$ is of ${\cal O}(M_Z)$.
For example, if $\mu(M_Z)$ is of ${\cal O}(1)$ TeV or larger, 
fine-tuning between $\mu^2(M_Z)$ and $m^2_{H_{u,d}}(M_Z)$ 
would be required to lead to the experimental value of $M_Z$ 
by using the relation (\ref{eq:mu-MZ}).
Following Ref.~\cite{Barbieri:1987fn}, 
we introduce the fine-tuning index,
\begin{eqnarray}
\Delta_{\mu} = \frac{1}{2}\left| \frac{\mu}{M_Z^2}
\frac{\partial M_Z^2}{\partial \mu} \right| =2\frac{\mu^2}{M_Z^2}.
\end{eqnarray}
When $\Delta_{\mu}=20 $ (10), the fine-tuning with the degree 
$5 \%$  ($10 \%$)
is required between $\mu^2$ and $m^2_{Hu,Hd}$ in order to lead to 
the experimental value of $M_Z$ in Eq.(\ref{eq:mu-MZ}).
Then, the conditions for $\Delta_{\mu} \leq 20 $ and 10 
correspond to $\mu(M_Z) \leq 280$ and 210 GeV, respectively.
The region with squares in figures corresponds to 
the parameter space with $\Delta_{\mu} \leq 20 $, 
where we also have $m_h \geq 114.4$ GeV and superparticle masses 
larger than experimental lower bounds. 
We refer such a region as a favored one.
Obviously, most of the favored regions correspond to the 
higgsino-like LSP region.
{}From these figures it is found 
that as $M_c$ increases, the favored region 
becomes narrow. If a larger value of $M_3(M_c)$ such as 500 GeV could
be chosen, the favored region also becomes narrow.

Finally, we give examples of soft masses, $\mu$, and
the lightest Higgs mass at $M_Z$ for cases  (i), (ii) and (iii) in
Tables \ref{tab1}, \ref{tab2} and \ref{tab3}. For comparison, the corresponding
values in the slepton LSP case (described by (blue) triangles in figures)
are also shown in the same tables. The examples corresponding to the
higgsino-like, wino-like, and slepton LSP cases are shown by A, B, and
C in the figures, respectively. It is found that there is not a
favored parameter space of the wino-like LSP in the case (ii) with $M_c=2\times10^{16}$ GeV because of an unsuccessful EWB.  
In the wino-like LSP region, typically the value of $r_1$ is large, 
that is, the bino mass is large compared with other gaugino masses.
Then, sleptons are heavier than squarks, and 
slepton masses are of ${\cal O}(1)$ TeV.
On the other hand, in the higgsino-like LSP region, 
the value $r_1$ is less than those values of $r_1$ in the wino-like LSP region.
Sleptons are lighter than squarks, but those spectra are compact 
compared with the case for the GUT relation. 
\begin{table}
\begin{center}
\begin{tabular}{|r||c|c|c|c|c|c|c|c|c|}
\hline
LSP~~~~~~~~ & \multicolumn{3}{|c|}{higgsino-like [A]} & \multicolumn{3}{|c|}{wino-like [B]} & \multicolumn{3}{|c|}{slepton [C]} \\
\hline
$M_c$ [GeV] & $2\times10^{16}$ & $10^{14}$ & $10^{12}$ & $2\times10^{16}$ & $10^{14}$ & $10^{12}$ & $2\times10^{16}$ & $10^{14}$ & $10^{12}$\\
\hline
$r_1$~~~~~~~~~ & 5 & 7 & 10 & 26 & 22& 19 & 2 & 2 & 10\\
\hline
$r_2$~~~~~~~~~ & 4 & 3.5 & 3.25 & 0.75 & 0.5 & 0.5 & 3.5 & 3.25 & 4 \\
\hline
$M_1(M_Z)$ [GeV]         & 617 & 773 & 992 &3210 & 2430 & 1890 & 247 & 221 & 992 \\
\hline
$M_2(M_Z)$ [GeV]         & 991 & 776 & 647 & 186 & 111 & 99.7 & 867 & 721 & 797 \\
\hline
$M_3(M_Z)$ [GeV]         & 865 & 774 & 695 & 865 & 774 & 695 & 865 & 774 & 695 \\
\hline
$m_{Q_{1,2}}(M_Z)$ [GeV] & 1140 & 900 & 744 & 932 & 740 & 623 & 1060 & 869 & 808 \\
\hline
$m_{u_{1,2}}(M_Z)$ [GeV] & 859 & 771 & 708 & 2170 & 1360 & 948 & 780 & 681 & 708 \\
\hline
$m_{d_{1,2}}(M_Z)$ [GeV] & 789 & 699 & 623 & 1270 & 896 & 698 & 769 & 675 & 623 \\
\hline
$m_{L_{1,2}}(M_Z)$ [GeV] & 895 & 655 & 528 & 1530 & 893 & 560 & 749 & 555 & 616 \\
\hline
$m_{e_{1,2}}(M_Z)$ [GeV] & 587 & 565 & 585 & 3050 & 1780 & 1110 & 235 & 162 & 585 \\
\hline
$\mu(M_Z)$ [GeV]         & 178 & 228 & 186 & 267 & 237 & 234 & 291 & 274 & 696 \\
\hline
$m_{h^0}$ [GeV]            & 116   & 116 & 115 & 116 & 118 & 116 & 115 & 114.5 & 115 \\
\hline
\end{tabular}
\end{center}
\caption{Soft masses, $\mu$, and the lightest Higgs mass at $M_Z$ in case (i)}
\label{tab1}
\end{table}
\begin{table}
\begin{center}
\begin{tabular}{|r||c|c|c|c|c|c|c|c|}
\hline
LSP~~~~~~~~ & \multicolumn{3}{|c|}{higgsino-like [A]} & \multicolumn{2}{|c|}{wino-like [B]} & \multicolumn{3}{|c|}{slepton [C]} \\
\hline
$M_c$ [GeV] & $2\times10^{16}$ & $10^{14}$ & $10^{12}$ & $10^{14}$ & $10^{12}$ & $2\times10^{16}$ & $10^{14}$ & $10^{12}$ \\
\hline
$r_1$~~~~~~~~~              & 7    & 7   & 12   & 26   & 24   & 2    & 5   & 10\\
\hline
$r_2$~~~~~~~~~              & 4.5  & 4   & 3.5  & 0.5  & 0.5  & 3.25 & 2.5 & 4.75 \\
\hline
$M_1(M_Z)$ [GeV]            & 864  & 773 & 1190 & 2870 & 2380 & 247  & 552 & 992 \\
\hline
$M_2(M_Z)$ [GeV]            & 1110 & 887 & 697  & 111  & 99.7 & 805  & 554 & 946 \\
\hline
$M_3(M_Z)$ [GeV]            & 865  & 774 & 695  & 774  & 695  & 865  & 774 & 695 \\
\hline
$m_{Q_{1,2}}(M_Z)$ [GeV]    & 1230 & 958 & 767  & 763  & 640  & 1030 & 797 & 879 \\
\hline
$m_{u_{1,2}}(M_Z)$ [GeV]    & 941  & 771 & 754  & 1550 & 1110 & 780  & 725 & 708 \\
\hline
$m_{d_{1,2}}(M_Z)$ [GeV]    & 812  & 699 & 636  & 971  & 754  & 769  & 686 & 623 \\
\hline
$m_{L_{1,2}}(M_Z)$ [GeV]    & 1040 & 732 & 590  & 1050 & 705  & 697  & 468 & 707 \\
\hline
$m_{e_{1,2}}(M_Z)$ [GeV]    & 822  & 565 & 702  & 2100 & 1400 & 235  & 404 & 585 \\
\hline
$\mu(M_Z)$ [GeV]            & 235  & 276 & 254  & 250  & 228  & 433  & 426 & 696 \\
\hline
$m_{h^0}$ [GeV]             & 116  & 115 & 115  & 119  & 117  & 116  & 115 & 115 \\
\hline
\end{tabular}
\end{center}
\caption{Soft masses, $\mu$, and the lightest Higgs mass at $M_Z$ in case (ii)}
\label{tab2}
\end{table}
\begin{table}
\begin{center}
\begin{tabular}{|r||c|c|c|c|c|c|c|c|c|}
\hline
LSP~~~~~~~~ & \multicolumn{3}{|c|}{higgsino-like [A]} & \multicolumn{3}{|c|}{wino-like [B]} & \multicolumn{3}{|c|}{slepton [C]} \\
\hline
$M_c$ [GeV] & $2\times10^{16}$ & $10^{14}$ & $10^{12}$ & $2\times10^{16}$ & $10^{14}$ & $10^{12}$ & $2\times10^{16}$ & $10^{14}$ & $10^{12}$\\
\hline
$r_1$~~~~~~~~~              & 5    & 7   & 13   & 28   & 22   & 17   & 2    & 2   & 10\\
\hline
$r_2$~~~~~~~~~              & 4    & 3.5 & 2.5  & 0.5 & 0.5  & 0.5  & 3.5   & 3.5 & 4 \\
\hline
$M_1(M_Z)$ [GeV]            & 617  & 773 & 1290 & 3460 & 2430 & 1690 & 247  & 221 & 992 \\
\hline
$M_2(M_Z)$ [GeV]            & 991  & 776 & 498  & 124  & 111  & 99.7 & 867  & 776 & 797 \\
\hline
$M_3(M_Z)$ [GeV]            & 865  & 774 & 695  & 865  & 774  & 695  & 865  & 774 & 695 \\
\hline
$m_{Q_{1,2}}(M_Z)$ [GeV]    & 1140 & 900 & 693  & 947  & 740  & 618  & 1060 & 896 & 808 \\
\hline
$m_{u_{1,2}}(M_Z)$ [GeV]    & 859  & 771 & 779  & 2320 & 1360 & 888  & 780  & 681 & 708 \\
\hline
$m_{d_{1,2}}(M_Z)$ [GeV]    & 789  & 699 & 643  & 1340 & 896  & 678  & 769  & 675 & 623 \\
\hline
$m_{L_{1,2}}(M_Z)$ [GeV]    & 895  & 655 & 509  & 1650 & 893  & 501  & 749  & 596 & 616 \\
\hline
$m_{e_{1,2}}(M_Z)$ [GeV]    & 587  & 565 & 760  & 3290 & 1780 & 994  & 235  & 162 & 585 \\
\hline
$\mu(M_Z)$ [GeV]            & 263  & 240 & 125  & 265  & 248  & 172  & 349  & 244 & 696 \\
\hline
$m_{h^0}$ [GeV]             & 121  & 121 & 121  & 124  & 123  & 121  & 121  & 120 & 120 \\
\hline
\end{tabular}
\end{center}
\caption{Soft masses, $\mu$, and the lightest Higgs mass at $M_Z$ in case (iii)}
\label{tab3}
\end{table}

The favored region with squares corresponds to $r_2\lesssim4$ and $r_1\lesssim
30$ and 
such values would be realized in several scenarios.
For example, the TeV-scale mirage mediation leads to 
$(r_1,r_2)=(7,3.5)$ \cite{Choi:2005hd}.
Furthermore, the generalized TeV-scale mirage mediation 
could lead to larger value of $r_1$ \cite{Abe:2005rx,Abe:2007je} 
such as $r_1={\cal O}(10)$.
As another example, when the F-term of ${\bf 75}$ in SU(5) GUT 
is dominant in the gaugino masses, 
we have $(r_1,r_2)=(5,3)$~\cite{Ellis:1984bm}.
On the other hand, the F-term of ${\bf 24}$ in SU(5) GUT 
leads to rather smaller values, i.e. $(r_1,r_2)=(0.5,1.5)$.
The minimal gauge messenger scenario also leads to 
small values, $(r_1,r_2)=(2.5,1.5)$ \cite{Dermisek:2006qj}.
Values of $r_1$ and $r_2$ depend on details of models in 
the moduli mediation scenario.
For example, the so-called O-II model leads to 
$(r_1,r_2)=(29/5,3)$ \cite{Brignole:1993dj,Horton:2009ed}.
In the general gauge mediation scenario, 
we would have naturally non-universal gaugino masses with 
$r_1,r_2={\cal O}(1)$, but their concrete values depend on 
details of models.

\subsection{Degeneracy}

One of important aspects in the superconformal flavor scenario 
is that squark and slepton  masses between the first and second 
families are degenerate.
It is important to study how much such degeneracy changes 
by varying $(r_1,r_2)$.
In this section, we study such an aspect.

\begin{figure}
\begin{center}
\begin{tabular}{c}
(a) $M_c=2\times10^{16}$ GeV \\
\includegraphics[scale = 1.18]{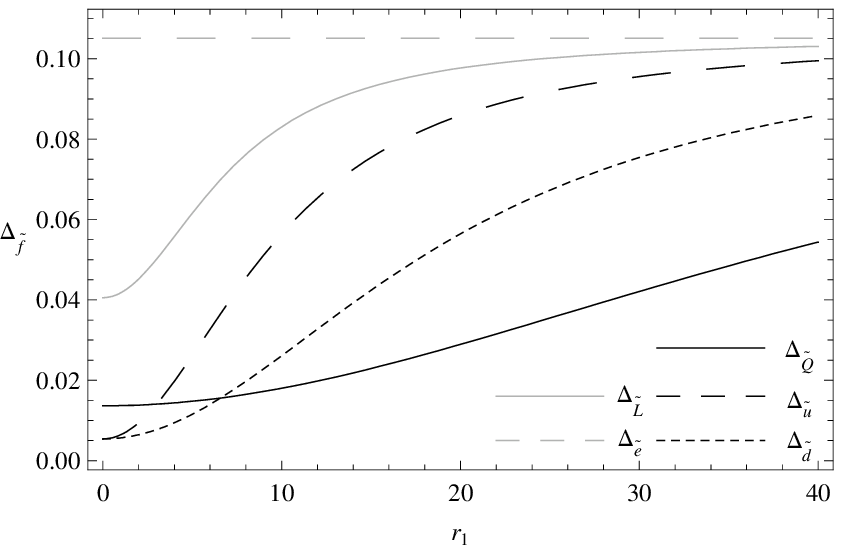} \\
(b) $M_c=10^{14}$ GeV \\
\includegraphics[scale = 1.18]{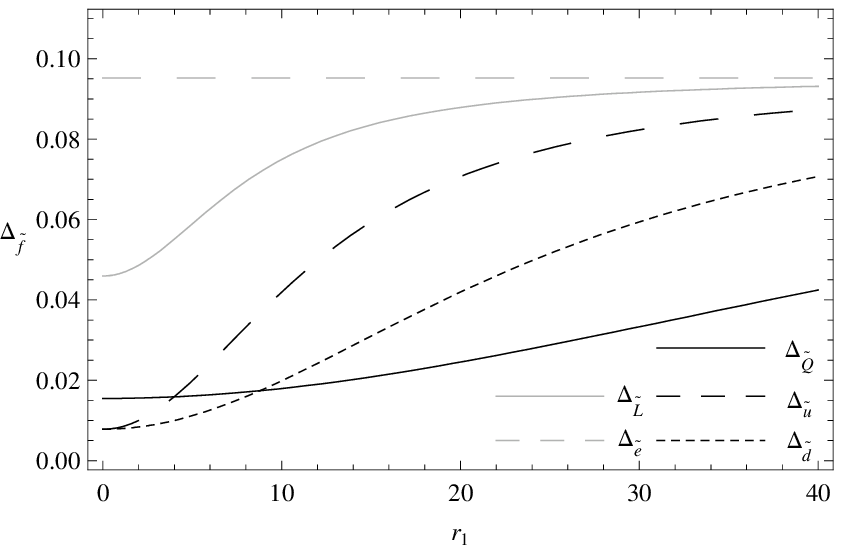} \\
(c) $M_c=10^{12}$ GeV \\
\includegraphics[scale = 1.18]{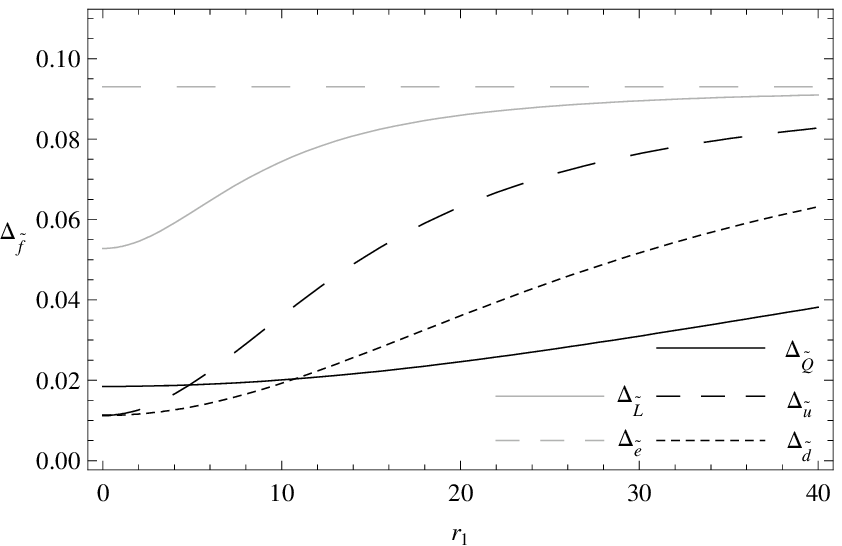}\\
\end{tabular}
\end{center}
\caption{Degeneracy degrees of soft scalar masses squared, $\Delta_{\tilde{f}}$
}
\label{fig4}
\end{figure}

As said in the previous section, soft scalar masses squared 
converge to flavor-dependent values of ${\cal O}(\alpha_aM_a^2)$ 
at $M_c$ as Eqs.~(\ref{eq:m-converge}) and (\ref{eq:m-converge-2}).
Those values are rather small and flavor-dependent, 
and flavor-independent  
radiative corrections between $M_c$ and $M_Z$ are 
dominant in values of soft scalar masses squared at $M_Z$.
Thus, we estimate the degeneracy degree of soft scalar masses 
squared by the following values,
\begin{eqnarray}\label{eq:degeneracy}
& & \Delta_{\tilde Q} = \Gamma_{Qi} \frac{m_{Qi}^2(M_c)}{m_{Qi}^2(M_Z)}, 
\qquad 
\Delta_{\tilde u} = \Gamma_{ui} \frac{m_{ui}^2(M_c)}{m_{ui}^2(M_Z)}, 
\qquad 
\Delta_{\tilde d} = \Gamma_{di} \frac{m_{di}^2(M_c)}{m_{di}^2(M_Z)}, 
\nonumber \\
& & \Delta_{\tilde L} = \Gamma_{Li} \frac{m_{Li}^2(M_c)}{m_{Li}^2(M_Z)}, 
\qquad
\Delta_{\tilde e} = \Gamma_{ei}
\frac{m_{ei}^2(M_c)}{m_{ei}^2(M_Z)}.
\end{eqnarray}
Recall that $m_{Qi}^2(M_c)$ is proportional to $1/\Gamma_{Qi}$ 
in Eq.~(\ref{eq:m-converge-2}) and 
$ \Gamma_{Qi}$ is flavor-dependent, where $ \Gamma_{Qi}$ would be 
of ${\cal O}(1)$.
In Eq.~(\ref{eq:degeneracy}), we have put the factors such as 
$ \Gamma_{Qi}$ in the right hand side in order to estimate 
the degeneracy degree for generic case.
Similarly, we have defined $\Delta_{\tilde u}$, $\Delta_{\tilde d}$, 
$\Delta_{\tilde L}$ and $\Delta_{\tilde e}$.
Note that these values depend on only $r_1$ and $r_2$.
Figures \ref{fig4} (a)$-$(c) 
show those values when we vary $r_1$ for $r_2=2$.
We obtain similar behaviors for other values of $r_2$ such as 
$r_2=3, 4$ or 5. 

First, we comment on the degeneracy of squark masses 
and left-handed slepton masses except for right-handed slepton masses.
For a large value of $r_1$ like $r_1\gtrsim 20$, 
the degeneracy becomes worse.
Thus, such a large value of $r_1$ would not be favorable from 
the viewpoint of the FCNC problems.
As $r_1$ increases, values of $\Delta_{\tilde Q}$, 
$\Delta_{\tilde u}$, $\Delta_{\tilde d}$, and $\Delta_{\tilde L}$
approach toward certain values.
That is because the bino mass $M_1$ becomes 
dominant in both the numerator and denominator 
of $\Delta_{\tilde Q}$, 
$\Delta_{\tilde u}$, $\Delta_{\tilde d}$, and $\Delta_{\tilde L}$ 
in the right hand side of Eq.~(\ref{eq:degeneracy}).
On the other hand, for a small value of $r_1={\cal O}(1)$, 
degeneracies of squark masses 
and left-handed slepton masses are still good.
For squarks, a value such as $r_1 \lesssim 10-20$ is still 
favorable. 
Thus, in these parameter regions, 
we could avoid the FCNC problem for $\Gamma_i \gtrsim 0.1$.

Next we comment on the right-handed slepton masses.
The degeneracy for 
right-handed slepton masses is weak.
We may face with the FCNC problem for $\Gamma_i ={\cal O}(0.1)$.
Indeed, the value of $\Delta_{\tilde e} $ does not change 
when we vary $(r_1,r_2)$, because 
both the numerator and denominator of $\Delta_{\tilde e} $ 
in the right hand side of Eq.~(\ref{eq:degeneracy}) 
depend on only the bino mass $M_1$.
Hence, the degeneracy degree of $\Delta_{\tilde e} $ can not 
be improved by the non-universal gaugino masses (\ref{eq:non-universal})
and the result on $\Delta_{\tilde e} $ is the same 
as one in Ref.~\cite{Kobayashi:2001kz,Nelson:2001mq}.
However, when we increase $r_1$ with $M_3(M_c)$ fixed, 
the bino and right-handed sleptons become heavy.
Such heavy right-handed sleptons would be helpful 
to improve the FCNC problem even for weak degeneracy like 
$\Delta_{\tilde e} ={\cal O}(0.1)$~\cite{FCNCbound}.
For example, most of the favored parameter space with 
the higgsino LSP and $\Delta_{\mu} \lesssim 20$, which is plotted by 
 squares, 
corresponds to $m_{\tilde e} > 500$ GeV.
In the wino LSP region, we have  $m_{\tilde e} ={\cal O}(1)$ TeV.

We have concentrated on soft scalar masses.
In general, A-terms also cause the FCNC problems.
Finally, we give a comment on A-terms.
Here, we consider the following A-terms 
\begin{eqnarray}
a^{ij}q_i q_j H,
\end{eqnarray}
which corresponds to the Yukawa couplings $y^{ij}q_iq_jH$ 
in the superpotential (\ref{eq:super-p}).
Such A-terms are also controlled by the superconformal dynamics.
Similar to the Yukawa couplings $y_{ij}$ in (\ref{eq:yukawa}), 
they behave as 
\begin{eqnarray}\label{eq:a-term}
a^{ij}(M_c) \sim a^{ij}(\Lambda) \left( \frac{M_c}{\Lambda}
\right)^{(\gamma_i +\gamma_j)/2},
\end{eqnarray}
and they are suppressed.
However, the ratio $A^{ij}=a^{ij}/y^{ij}$ is important in 
the FCNC problem, but such a ratio is not suppressed 
during the conformal regime, because the 
renormalization group flows of $y^{ij}$ and $a^{ij}$ 
are the same.
In general, flavor-dependent values of $A^{ij}$  
are strongly constrained by FCNC experiments.
In particular, there is a strong constraint on 
the $\mu \rightarrow e \gamma$ process.
At any rate, the ratio $A^{ij}=a^{ij}/y^{ij}$ can not be 
controlled by the superconformal dynamics.
However, large slepton masses such as 500 GeV or more, 
which are shown in Tables \ref{tab1}, \ref{tab2} and \ref{tab3}, 
are favorable to avoid the FCNC problem through 
the A-terms~\cite{FCNCbound}.

\section{Conclusion and discussion}

We have studied on superparticle spectra in the 
superconformal flavor scenario.
We have assumed the non-universal gaugino masses.
The non-universality in the gaugino masses is 
quite important.
The slepton LSP is not evadable with the GUT relation, 
unless the $U(1)_Y$ D-term is sizable.
However, the non-universality of gaugino masses 
can lead to the wino-like or higgsino-like LSP.
Furthermore, such a parameter space 
includes the region, where the fine-tuning between $\mu$ 
and Higgs soft masses  
could be improved/ameliorated.
The degeneracy of soft scalar masses squared 
does not change drastically by varying ratios of 
gaugino masses for the region $r_1,r_2={\cal O}(1)$.
The degeneracy of scalar masses for squarks and left-handed 
sleptons would be good enough to avoid the FCNC problem, 
but the degeneracy of right-handed slepton masses is 
weak.
However, the overall size of right-handed slepton masses 
becomes larger when the bino becomes heavier.
That would be favorable to avoid the FCNC problem 
due to soft scalar masses and A-terms.
When $r_1 \gtrsim 20$, the degeneracy of squark 
and slepton masses becomes weak.

Our analysis shows that the superconformal flavor scenario 
with non-universal gaugino masses is quite interesting.
Thus, it is intriguing to study model construction leading to 
realistic Yukawa couplings, although explicit model 
construction is still a challenging issue.
Alternatively, from the AdS/CFT viewpoint, 
it would be important to study model building in
the 5D warped background \footnote{See e.g. \cite{Choi:2003di} 
and references therein.}.

\subsection*{Acknowledgement}

T.~K. is supported in part by the Grant-in-Aid for 
Scientific Research No.~20540266 from the 
Ministry of Education, Culture, Sports, Science and Technology of Japan.
T.~K. and Y.~N. are also supported in part 
by the Grant-in-Aid for the Global COE 
Program "The Next Generation of Physics, Spun from Universality and 
Emergence" from the Ministry of Education, Culture, Sports, Science and 
Technology of Japan. The work of R. T. is supported by the DFG-SFB TR 27.
Y.~N. is grateful to Institute for Advanced Study for their hospitality where part of this work was done.


\end{document}